\documentclass[10pt,
a4paper,
twocolumn
]{article}
\RequirePackage[labelsep=space,tableposition=top,font=small]{caption}
\usepackage{subcaption}
\usepackage{booktabs} 
\usepackage{multirow}
\usepackage{siunitx} 
\usepackage{enumitem}
\usepackage{appendix}

\usepackage[left=2cm,right=2cm,top=2cm,bottom=2cm]{geometry}
\usepackage{amsmath}
\usepackage{amssymb}
\usepackage{braket}
\usepackage{mathtools}
\usepackage{graphicx}
\usepackage{verbatim}
\usepackage{textcomp}
\usepackage{url}
\usepackage{cprotect}
\usepackage{tikz}
\usepackage{color}
\usepackage{pgfgantt}
\usepackage{pdflscape}
\usepackage{slashed}
\usepackage{caption}
\usepackage{subcaption}
\usepackage{floatrow}

\newcommand{\bfk}{\mathbf{k}}
\newcommand{\bfq}{\mathbf{q}}

\newcommand{\vect}[1]{\mathbf{#1}}
\newcommand{\create}[1]{c^{\dagger}_{#1, \sigma}}
\newcommand{\crup}[1]{c^{\dagger}_{#1, \uparrow}}
\newcommand{\crdo}[1]{c^{\dagger}_{#1, \downarrow}}
\newcommand{\annihi}[1]{c^{}_{#1, \sigma}}
\newcommand{\anup}[1]{c^{}_{#1, \uparrow}}
\newcommand{\ando}[1]{c^{}_{#1, \downarrow}}

\newcommand{\psit}{\Psi_\mathrm{T}}

\newcommand{\nup}{N_\uparrow}
\newcommand{\ndo}{N_\downarrow}

\newcommand{\EF}[0]{E_\mathrm{F}}

\newcommand{\kF}[0]{k_\mathrm{F}}

\newcommand{\abs}[1]{\lvert #1 \rvert}
\newcommand{\figref}[1]{Fig.~\ref{#1}}

\newcommand{\nuup}[0]{\nu_{\uparrow}}
\newcommand{\nudn}[0]{\nu_{\downarrow}}

\newcommand{\reff}[0]{r_\mathrm{e}}
\newcommand{\rs}[0]{r_\mathrm{s}}
\newcommand{\cutlen}[0]{L_\mathrm{c}}

\definecolor{pu}{RGB}{200,50,200}
\definecolor{gr}{RGB}{0,150,0}
\definecolor{bl}{RGB}{0,120,255}
\definecolor{re}{RGB}{255,0,0}

\begin{document}
\title{Diffusion Monte Carlo study of a spin-imbalanced two-dimensional Fermi gas with attractive interactions}
\author{D.C.W.~Foo$^1$ \and G.J.~Conduit$^1$}
\date{$^1$\emph{Cavendish Laboratory, J.J. Thomson Avenue, Cambridge, CB3 0HE, 
United Kingdom}\\[2ex]
\today}

\twocolumn[
  \begin{@twocolumnfalse}
    \maketitle
    \begin{abstract}
We probe the superconducting gap in the zero temperature ground state of an attractively interacting spin-imbalanced two-dimensional Fermi gas with Diffusion Monte Carlo. A condensate fraction at nonzero pair momentum evidences a spatially non-uniform superconducting order parameter. Comparison with exact diagonalisation studies confirms that the nonzero condensate fraction across a range of nonzero fermion pair momenta is consistent with non-exclusive pairing between majority and minority fermions, an extension beyond FFLO theory. 
    \end{abstract}
  \end{@twocolumnfalse}
]

\section{Introduction}\label{sec:intro}

One of the successes of condensed matter physics is the description of the phenomenon of superconductivity by Bardeen, Cooper, and Schrieffer (BCS)~\cite{bcs} which described Cooper pair (an up and down-spin fermion) formation close to the Fermi surface in a many-body context. While only applicable for a certain class of systems, specifically those with an equal number of spin-up and spin-down fermions, BCS theory well describes many superconducting materials ancient~\cite{supconelemrev,supconcomprev} and modern~\cite{consupconquacrys,consupconhight} and so naturally numerous extensions have been proposed and considered. 

For example, for systems under the influence of a strong magnetic field, a partial alignment of the fermion spins leads to a population imbalance and a shift in the sizes of the Fermi surfaces. While in the BCS theory such a strong magnetic field would suppress superconductivity entirely~\cite{bcs}, numerous proposals and extensions have arisen since then that might allow for such exotic pairing. For example, the minority spin species could pay the kinetic energy cost to promote fermions up to the Fermi level of the majority species, breaching the so-called Chandrasekhar-Clogston limit~\cite{chandrasekhar_limit,clogston_limit}, also referred to as the Pauli limit. Alternatively, pairing might form at the minority species Fermi level, leaving the fermions of the majority species above unpaired as in breached superconductivity~\cite{breachsupercon1,breachsupercon2,breachsupercon3}. Yet another possibility was proposed by Fulde and Ferrell (FF)~\cite{ffloff}, and Larkin and Ovchinnikov (LO)~\cite{fflolo} where fermions remain at their respective Fermi levels and then pair from opposite sides, resulting in a Cooper pair with net momentum. The superconducting gap then oscillates at the concomitant wavevector with the FF phase having a single wavevector and the LO phase two equal but opposite wavevectors. 

Recent work extended the idea of pairing at nonzero net momentum further with the introduction of the communal pairing state~\cite{multipartsupconfew,commupair} where the superconducting gap has peaks at multiple nonzero momenta. The key distinction between communal pairing theory and the FFLO family of pair density wave theories is that communal pairing theory by construction heavily features nonexclusive pairing between fermions at the Fermi surfaces of the minority and majority spin-species. This allows all of the fermions to participate in pairing, reducing the overall energy of the system through the contribution of correlation energy, compared to FFLO where pairing is one-to-one so not all the fermions on the majority spin-species are involved, presenting an opportunity to variationally include additional fermion states. Therefore, the optimal ratio of majority to minority spin fermions in the communal pairing phase is naturally predicted to be the ratio of the densities of states in momentum space at the Fermi surface, $\nup/\ndo=\nuup/\nudn$~\cite{commupair}.

While the formulation of BCS theory was preceded by the experimental observations of Onnes in 1911~\cite{Onnes1,Onnes2}, observation of spatially non-uniform pairing superconducting states in spin-imbalanced systems remains an experimental challenge despite considerable effort across a wide range of physical systems, including heavy fermion systems~\cite{ffloexpthf}, iron-based superconductors~\cite{ffloexptfe1,ffloexptfe2,ffloexptfe3,ffloexptfe4}, asymmetric $d$-wave superfluids~\cite{ffloexptdw}, layered organic superconductors~\cite{ffloexptorg1,ffloexptorg2,ffloexptorg3,ffloexptorg4}, layered superconductor-ferromagnet hybrid structures~\cite{ffloexptsflayer}, quasicrystals~\cite{ffloquasicrys} and ultracold atomic gases~\cite{ffloexptopt1,ffloexptopt2}. 

Organic superconductors in particular have been a key system of interest, emerging as the leading candidate for observation of spatially modulated superconductivity owing to their crystals growing relatively cleanly and thus granting the superconducting pairs a long mean free path compared to their coherence length~\cite{ffloorg}, their high degree of customizability through the attaching of various functional groups in addition to doping, and their inherently quasi-2D structure, as low dimensionality is expected to enhance FFLO physics~\cite{lowd}. However, recent experimental developments in the field of ultracold atomic gases~\cite{uniformtrap} promise new routes to realization of and deep insight into the spatially non-uniform pairing state, with the technique already being used to probe Fermi gases with and without a strong spin-imbalance~\cite{trapfg1,trapfg2}. The precise level of control and ability to impose spin-imbalance without an applied magnetic field that may disrupt superconductivity make ultracold atomic gases a particularly atractive system. 

Another recent material that shows potential is the lanthanum superhydrides~\cite{lah}, conventional superconductors with high critical temperature and thus high critical field which may support superconducting states even with a spin-imbalance and thereby manifest spatially non-uniform pairing. Beyond terrestrial experiments, theories of more exotic matter such as neutron superfluids in the crust of magnetars~\cite{magnetar} and quark matter~\cite{quark} also support the existence of superconducting states with spatially non-uniform pairing, and indeed the high energy physics and quantum chromodynamics communities have long known of and been searching for such states~\cite{casalbuoni}.

Furthermore, the ongoing increase in computing speed and power has made it possible to simulate many-body quantum systems, affording us a fresh avenue of investigation into the pairing structure of these exotic systems. In the present study we focus exclusively on two-dimensional systems as low dimensionality is thought to enhance the stability of FFLO-like phases~\cite{fflostability1,fflostability2,fflostability3} and is the lowest dimension in which states different from FFLO but with space varying gap parameter are predicted~\cite{multipartsupconfew}.

Diffusion Monte Carlo (DMC) simulations offer a fast method to study spin-imbalanced fermion gases~\cite{riospairingorb,dmcfermion1,dmcfermion2}. It is exact except for a fixed node approximation and so includes all orders of correlators and loop diagrams. Furthermore, the temperature of the system can be kept constant, even at absolute zero, and so the fermion spins can be prevented from relaxing. Using this technique, we investigate fermion pairing at nonzero pair momenta, demonstrating a spatially non-uniform superconductor. Furthermore, the distribution of the condensate fraction in momentum space is similar to that predicted by communal pairing theory~\cite{commupair} and markedly different to that predicted by FFLO theory. This provides evidence that non-exclusive pairing is the ground state of the spin-imbalanced superconductor.

In the next section we introduce the Hamiltonian for the problem and summarise the numerical methods employed for this study, namely Variational Monte Carlo (VMC) and DMC. Results obtained for both the spin-balanced and spin-imbalanced cases are then presented in Section \ref{sec:results}, followed by a discussion on the effects of changing various simulation parameters. Conclusions are presented in Section \ref{sec:conclusion}.

\section{Quantum Monte Carlo}
We use the \textsc{casino} code~\cite{casino} to perform our quantum Monte Carlo study of the Hamiltonian
\begin{align}
H = \sum_{i}\frac{\nabla^{2}_{i}}{2}+\sum_{i,j}V(\vect{r}_i-\vect{r}_j)
\end{align}
where $i$ and $j$ index the fermions, $\vect{r}_i$ is the position vector of fermion $i$, and $V$ is the interaction potential. The fermions are of equal mass and we work in a combination of natural and Hartree units so $\hbar=c=e=m=1$.

\subsection{Pseudopotential}\label{sec:pseudopot}
We use an ultratransferable pseudopotential (UTP)~\cite{2dscapseudopot}, a continuous, differentiable, piecewise defined polynomial that can be optimized to match the scattering phase shift of a known target potential over a range of incident momenta. The UTP is defined such that it is nonzero for distances less than a cutoff length $\cutlen$ and zero beyond. $\cutlen$ thus controls the extent of the potential in real space and in App.~\ref{subsec:simparams} is chosen to be equal to $\rs$, the average interparticle separation and the typical length scale above which a fermion could erroneously feel a potential from two other fermions simultaneously. The scattering length was chosen to be such that the superconducting coherence length was less than the simulation cell size and the effective range was fixed at zero.

\subsection{Trial wavefunction}\label{subsec:trialwf}
We follow after previous work ~\cite{riospairingorb} and employ a Slater-Jastrow trial wavefunction of the form $\psit = \mathrm{e}^{-J}\det[\phi(\vect{s}_{i,j})]$. The determinant ensures the correct fermionic spin-symmetry. The pairing orbital $\phi(\vect{s}_{i,j})$ comprises plane wave and polynomial expansions in the fermion separations $\vect{s}_{i,j}$. The plane wave part of the pairing wavefunction equals the exact Hartree-Fock ground state wavefunction for the non-interacting fermion gas, and in the presence of attractive interactions, the polynomial component of the pairing orbital can shift the nodal surface to smoothly transform to a superconducting wavefunction. To capture additional fermion correlations, we include the Jastrow factor $J$ which is a function of all opposite-spin fermion separations comprising a short range isotropic $u$ term, anisotropic $p$ terms~\cite{drumjas}, and a $\nu$ term~\cite{nujas} that reflects the simulation cell symmetry.

\subsection{Monte Carlo and expectation values}\label{sec:vmcdmc}
VMC is used to optimise the trial wavefunction by finding the minimum energy with respect to its variational parameters. The trial wavefunction optimised by VMC was the starting point for DMC~\cite{casino, qmcmethods, qmcreview}, which treats the Schr\"{o}dinger equation as a diffusion equation in imaginary time and evolves the wavefunction to project out our best estimate of the ground state~\cite{dmcintro}. 

We probe the superconducting state by measuring the expectation values of the momentum density and the condensate fraction. The momentum density is the fourier transform of the one-body density matrix and is defined as $n_{\bfk,\sigma} \equiv \langle\create{\bfk}\annihi{\bfk}\rangle$. The condensate fraction is a modified form of the two-body density matrix and is defined explicitly as $f_{\bfq} \equiv \sum_\bfk (\langle\crup{\bfk}\crdo{\bfq-\bfk}\ando{\bfq-\bfk}\anup{\bfk}\rangle - n_{\bfk,\uparrow}n_{\bfq-\bfk,\downarrow})$. For the BCS wavefunction, it evaluates to $f_{\bfq} = \delta_{\bfq,\vect{0}}\tfrac{\abs{\Delta}}{8\pi}(\tan^{-1}\tfrac{\mu}{\abs{\Delta}}+\tfrac{\pi}{2})\to\delta_{\bfq,\vect{0}}\tfrac{\abs{\Delta}}{8}$ for $\mu\gg\abs{\Delta}$, providing an estimate of the superconducting gap $\abs{\Delta}$. 

\subsection{Simulation setup and convergence}\label{sec:finitesizeerror}
With the simulation methodology in place the final step is to set up the system. The major considerations are the size and shape of the simulation cell that could lead to finite size errors. Discussion of other simulation parameters including scattering length, pseudopotential cutoff length, DMC timestep, and number of DMC walkers is deferred to App.~\ref{subsec:simparams}.

We employed two forms of simulation cell; a rhomboidal and square box, giving triangular and square tilings respectively in momentum space. Both geometries gave quantitatively similar results. A triangular lattice has the densest possible tiling of momentum points in 2D, giving the closest to circular Fermi surfaces and thereby minimizing finite size effects~\cite{finitesizeeffect,simdark}. This was confirmed by varying the number of particles simulated in our DMC studies varied from 26 to 164, which compares favourably with DMC studies conducted on other systems~\cite{dmcN1,dmcN2}. Finally, in the non-interacting and balanced system limits the results obtained compared favourably to known analytical results in the thermodynamic limit. Therefore, in these paradigmatic systems our simulations were free of finite size effects.

\section{Results}\label{sec:results}
\subsection{Spin-balanced BCS state}\label{subsec:bcscf}

\begin{figure}[t]
\centering
\includegraphics[width=0.9\linewidth]{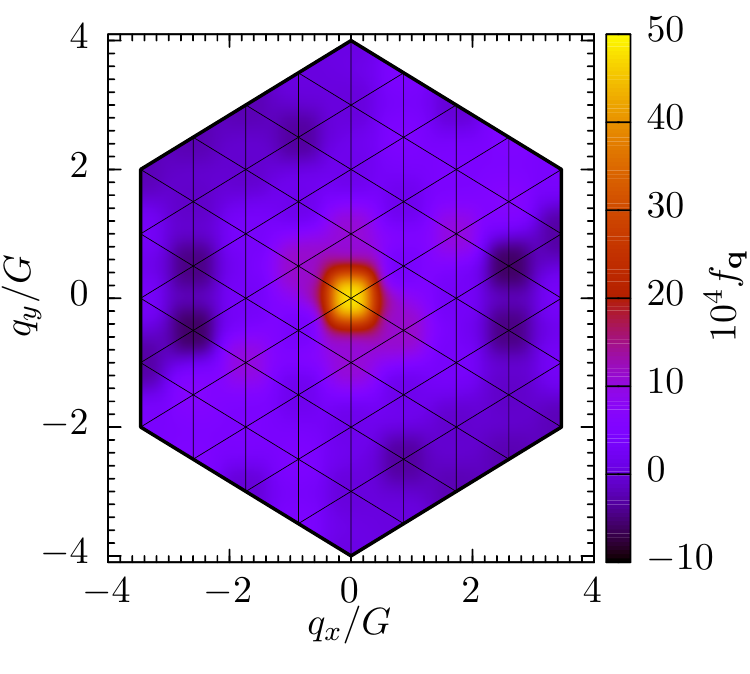}
\caption{Plot of the condensate fraction in pair-momentum space for the spin-balanced case with 37 fermions of each species. The wavevectors are scaled in units of the reciprocal lattice vector $G$ and black lines denote the $q$-space grid.}
\label{bcscfplot}
\end{figure}

We start from the well-established spin-balanced BCS system to confirm the accuracy of our simulations, and later explore imbalance. To build our investigation from a solid platform, we first study a spin-balanced system with 37 spin up and spin down fermions. We select a scattering length $a=5.6\rs$ and effective range $\reff=0$ to ensure that the superconducting coherence length is less than the size of the simulation cell. 

The accumulated condensate fraction is shown in Fig.~\ref{bcscfplot}. The condensate fraction at $q=0$ was 8 sample standard deviations above zero while those at every other $q$ point were within 2 sample standard deviations of zero. The reduction in energy of the interacting system compared to the non-interacting system meanwhile was 0.54 $\EF$ which agrees with that obtained from analytic calculation~\cite{2dscaverhaar84,fermigasreview} and other numerical studies~\cite{shi}. The results obtained for the spin-balanced case are therefore in line with theoretical expectations of BCS theory~\cite{bcs} and we proceed with confidence in the veracity of the simulations.

We note for completeness that the condensate fraction was also gathered for pairs of the same spin-species to confirm the presence or absence of induced p-wave superfluidity~\cite{fflorevision3}. The values of the intra-spin condensate fraction were more than 10 orders of magnitude smaller than those for the inter-spin condensate fraction and were indistinguishable from zero at all pair momenta for both the spin-balanced case presented above and the spin-imbalanced cases discussed below.

\subsection{Spin-imbalanced superconducting state}\label{subsec:basecase}

Having confirmed the accuracy of DMC simulations in the spin-balanced case, we now turn to the simplest class of spin-imbalanced systems with a 2:1 ratio of states on the Fermi surfaces, and so communal pairing theory predicts a $(\nup,\ndo) = (2,1)$ communal pairing instability while FFLO theory predicts pairs with nonzero net momentum. We conduct our study on the triangular lattice with 61 spin-up and 19 spin-down fermions, and in the square lattice with 25 spin-up and 9 spin-down fermions. On both lattices we use $a=6.0\rs$ and $\reff=0$.

\begin{figure}[t]
\centering
\includegraphics[width=0.9\linewidth]{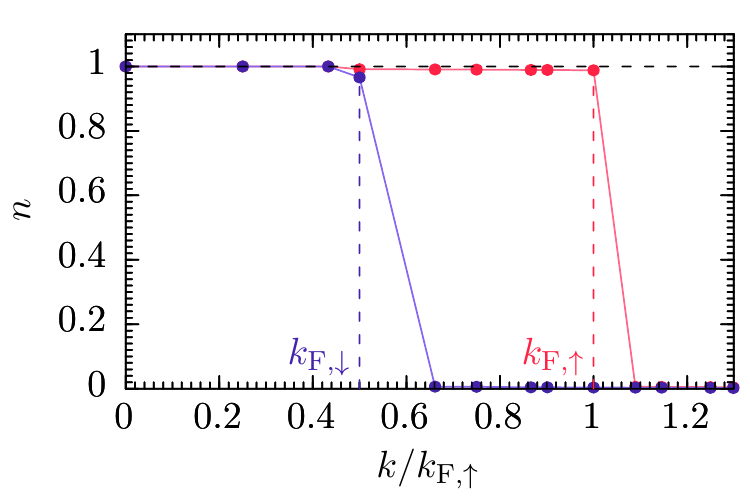}
\caption{Momentum density $n$ of the majority up (red) and minority down (blue) spin-species on the triangular lattice. The vertical dashed lines indicate the respective Fermi momenta and the the horizontal black dashed line denotes $n=1$. The finite slope at the Fermi momenta are due to the finite resolution of the momentum space lattice.}
\label{momden}
\end{figure}

\subsubsection{Momentum density}
We first examine the momentum density, with the results on the triangular lattice in Fig.~\ref{momden}. Both spin-species have momentum density close to unity beneath their respective Fermi momenta and close to zero above. A breached superfluid~\cite{breachsupercon1,breachsupercon2,breachsupercon3} would have the majority species exhibit depletion at the minority species Fermi momentum and a system crossing the Chandrasekhar-Clogston limit~\cite{chandrasekhar_limit,clogston_limit} would have finite momentum density of the minority species at the majority species Fermi momentum, so it is clear that the system has not relaxed into either of those possible states. Knowing this, we can now move on to study the emergence of superconductivity by examining the condensate fraction.


\begin{figure}[t]
\centering
\begin{subfigure}{0.85\textwidth}
 \centering
 \includegraphics[width=\linewidth]{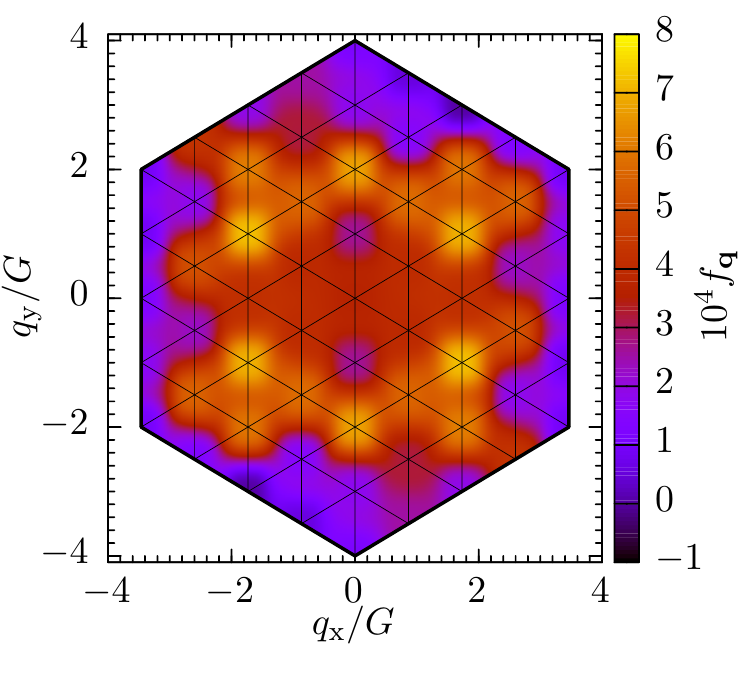}
 \caption{Triangular lattice}
 \label{2_1tri}
\end{subfigure}
\\
\begin{subfigure}{0.85\textwidth}
 \centering
 \includegraphics[width=\linewidth]{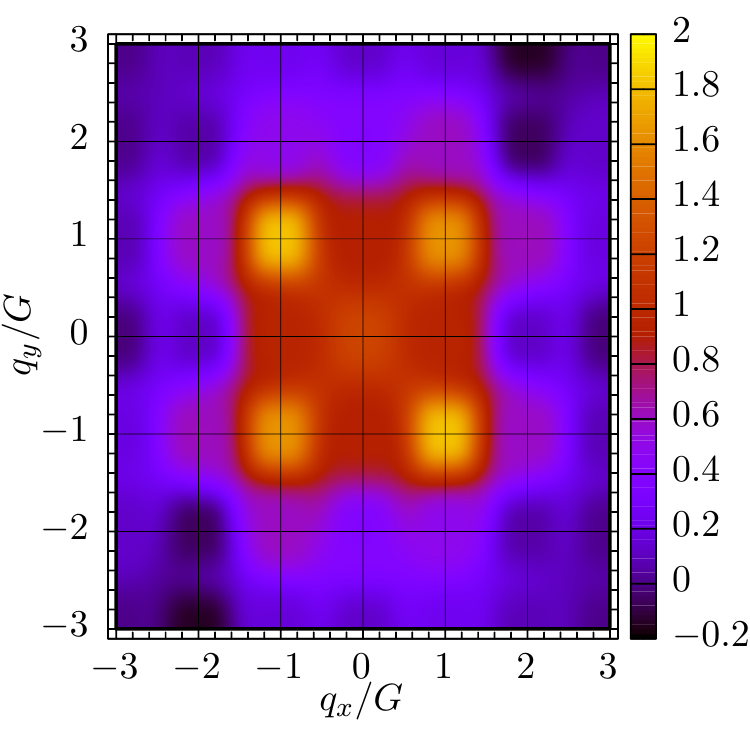}
 \caption{Square lattice}
 \label{2_1sq}
\end{subfigure}
\caption{Contour plot of the condensate fraction in momentum space for the spin-imbalanced case with 61 and 19 fermions of the majority and minority species respectively on the triangular lattice and 25 and 9 on the square lattice.}
\label{2_1}
\end{figure}

\subsubsection{Condensate fraction}
The condensate fraction for the spin-imbalanced system with 61 majority and 19 minority species fermions is shown in Fig.~\ref{2_1tri}. Six major peaks in the condensate fraction are visible at the points 2 $G$ units away from the origin, where $G$ is the magnitude of the reciprocal lattice vector. This is the first observation in a first principles simulation of pairing at finite momentum in two or more dimensions and so could be the first numerical evidence of a FFLO or other exotic spatially modulated pairing phase, but not the BCS phase. The result is qualitatively consistent with the spatially modulated pairing phase observed in experimental~\cite{ffloexptopt2} and numerical studies of one-dimensional systems~\cite{fflo1d}, and with few-particle studies~\cite{multipartsupconfew,fewpart}. We therefore now proceed to characterise the pairing to understand the correlations in the ground state.

The condensate fraction exhibits the six-fold rotational symmetry of the underlying momentum space lattice, in agreement with low temperature studies of spin-imbalanced pairing~\cite{fflorevision1,fflorevision2} that predicts an increase in the number of pairing momenta, $\vect{q}$, in the ground state. However, a key characteristic of the DMC results in \figref{2_1tri} is that they show statistically significant pairing at several momenta, $q<4G$, that are not at the optimal magnitude predicted by FFLO theory, and decays radially. This is a significant departure as the family of FFLO theories \cite{ffloff,fflolo,fflorevision1,fflorevision2} predicts a single optimal magnitude of pairing momenta and zero pairing amplitude otherwise.

Similar results are seen in Fig.~\ref{2_1sq} where 25 majority and 9 minority fermions have been placed in a square lattice. The condensate fraction reflects the rotational symmetry of the underlying momentum space lattice, a feature shared with crystalline FFLO theories~\cite{fflorevision1} and is nonzero beyond that of the optimal pairing momenta predicted by FFLO theory. While nonzero pairing at nonoptimal $q$ is not present in FFLO theory or any of its derivatives, it is however consistent with communal pairing~\cite{commupair}.

\subsubsection{Characterisation of the communal state}
\begin{figure}[t]
	\centering
	\includegraphics[width=0.9\linewidth]{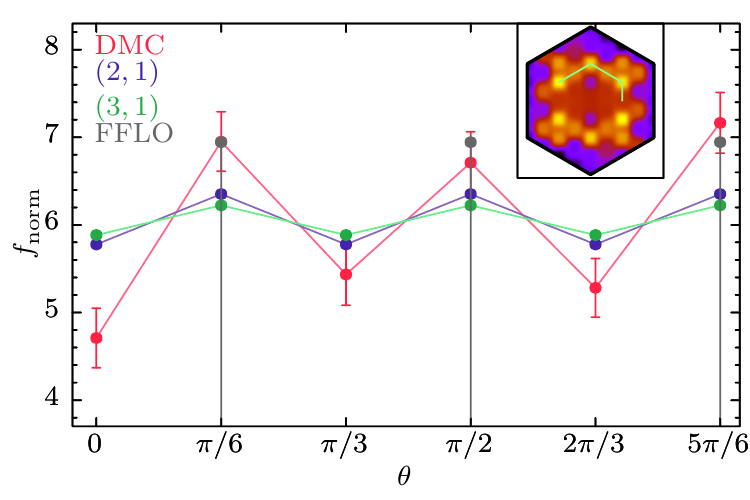}
	\caption{Plot of normalised condensate fraction $f^{}_{\mathrm{norm}}$ for states composed of pairs near the respective Fermi surfaces obtained using DMC (red) and predicted using exact diagonalisation assuming $(\nup,\ndo)=(2,1)$ (blue), $(\nup,\ndo)=(3,1)$ (green) and normal FFLO pairing with $(\nup,\ndo)=(1,1)$ (grey) against angle of the pair momentum vector. Inset: a copy of \figref{2_1tri} with a bright green curve indicating the displayed states of the plot.}
	\label{cfractan}
\end{figure}

\begin{figure}[t]
	\centering
	\includegraphics[width=0.9\linewidth]{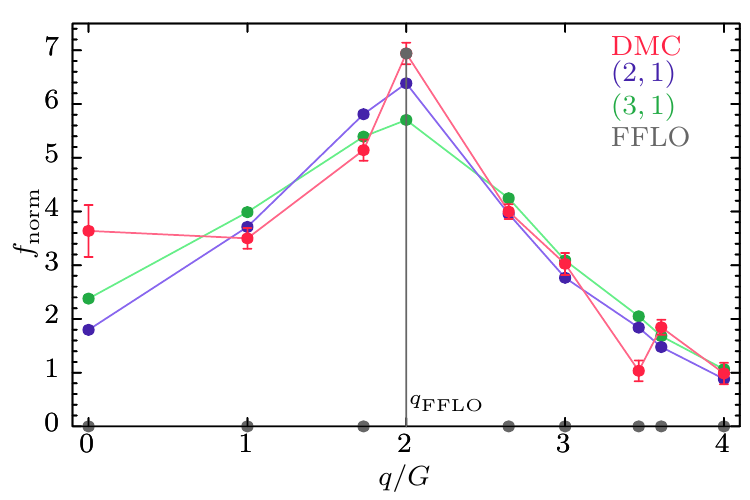}
	\caption{Plot of normalised condensate fraction $f^{}_{\mathrm{norm}}$ obtained using DMC averaged over $q$ (red) and condensate fraction predicted using exact diagonalisation of the states beneath the respective Fermi surfaces assuming $(\nup,\ndo)=(2,1)$ (blue), $(\nup,\ndo)=(3,1)$ (green) and normal FFLO pairing with $(\nup,\ndo)=(1,1)$ (grey) against $q$. The dashed black line marks out $q=q^{}_{\mathrm{FFLO}}$, the optimal magnitude of pairing momentum as predicted by FFLO theory and the only point on the FFLO curve where $f^{}_{\mathrm{norm}}\neq0$.}
	\label{cfraccomp}
\end{figure}

The condensate fraction indicates that the superconducting correlations are consistent with communal pairing. To probe the nature and number of fermions in the communal pairing state, we follow the prescription of Ref.~\cite{multipartsupconfew} and perform exact diagonalisation focusing on $(\nup,\ndo)=(2,1)$ or $(3,1)$ fermions in a subset of the momentum states used in the DMC study, specifically those at the Fermi surfaces of the respective spin-species, and calculate the condensate fraction averaged across pair momenta of fixed $q$ as a function of $q$. The strength of the contact interaction for the exact diagonalisation study was chosen to match that used in the DMC study. Results are shown in Fig.~\ref{cfractan} in the azimuthal direction and Fig.~\ref{cfraccomp} radially, along with the averaged results obtained using DMC.

The red curve in \figref{cfractan} shows the condensate fraction obtained from DMC at the values of pair momentum indicated by the bright green line in the inset. These values were chosen as they are the ones which involve pairing of fermions at or near their respective Fermi momenta. The red curve in \figref{cfraccomp} shows the angle-averaged condensate fraction obtained from DMC, where the average is taken over all pair momenta of equal magnitude.

The grey lines show the condensate fraction obtained when only one up and one down-spin fermion is allowed, as in the family of FFLO theories. FF theory predicts a single peak at a particular $\bfq$~\cite{ffloff}, LO theory predicts two peaks at $\bfq$ and $-\bfq$~\cite{fflolo}, and crystalline FFLO theory predicts multiple peaks for all $\lvert\bfq\rvert=q^{}_{\mathrm{FFLO}}$~\cite{fflorevision1,fflorevision2}. Our results confirm that having pre-selected for a single up and down-spin fermion, the crystalline FFLO ground state is the most stable out of these, in line with previous results~\cite{fflorevision1,fflorevision2}, with the condensate fraction equally shared by all symmetry related pair momentum vectors at this magnitude, as seen in \figref{cfractan}. In the specific system studied here, $q^{}_{\mathrm{FFLO}}=2G$, as seen in Fig.~\ref{cfraccomp}. To make a fair comparison between the DMC and exact diagonalisation results, we have normalized the condensate fraction obtained from exact diagonalisation so that the weighted squared deviation from the DMC results is minimized.

If instead communal pairing is considered, the results obtained from exact diagonalisation of both $(\nup,\ndo)=(2,1)$ and $(3,1)$ are quantitatively similar to those obtained from DMC, with both sets of results exhibiting three key features. First and foremost, both have a nonzero condensate fraction at many values of $q$ including $q=0$, an essential feature of communal pairing theory that is in contrast to the predictions of FFLO theory. This is a direct consequence of considering non-exclusive communal pairing. Secondly, both DMC and communal pairing have a global maximum at $q=q^{}_{\mathrm{FFLO}}$ as this corresponds to the paired fermions being at their respective Fermi levels and thereby minimising their kinetic energy. Finally, both DMC and communal pairing curves exhibit a decay in the condensate fraction for $q>q^{}_{\mathrm{FFLO}}$ which is due to the increasing kinetic energy cost of the fermion pairs. 

The quality of agreement between the communal pairing exact diagonalisation results and the DMC data can be quantified by the ratio of the weighted sum of squared deviations of the DMC results from either set of exact results in Fig.~\ref{cfractan} (azimuthal) or Fig.~\ref{cfraccomp} (radial), where the weights are the sample variances of the DMC data. This test statistic shows that the DMC results obtained are 27 times better described by an underlying $(\nup,\ndo)=(3,1)$ communal state and 38 times better described by an underlying $(\nup,\ndo)=(2,1)$ communal state than by FFLO pairing. This provides strong evidence that the state observed in DMC is not only communal, but has the appropriate values of $(\nup,\ndo)=(2,1)$. 

Similar results were obtained on performing exact diagonalisation at the Fermi surface of the system with 25 and 9 fermions on the square lattice; FFLO theory predicts a 4-fold degenerate peak at $q^{}_{\mathrm{FFLO}}=\sqrt{2}G$ and zero condensate fraction otherwise while the communal exact diagonalisation results for $(\nup,\ndo)=(2,1)$ and $(3,1)$ exhibited nonzero condensate over a range of momenta with a global maximum at $q^{}_{\mathrm{FFLO}}$. The test statistic obtained repeats the conclusion that the system is best described by a communal state with $(\nup,\ndo)=(2,1)$.

The mismatch between the DMC and communal exact diagonalisation results, particularly at $q=0$, may be due to a number of factors. Firstly, exact diagonalisation only accounts for a subset of the allowed momentum states without considering states above the Fermi surfaces, and secondly, that exact diagonalisation was carried out for only 2 (FFLO), 3 or 4 (communal) particles in total.

\begin{figure}[t]
\centering
\begin{subfigure}{0.85\textwidth}
 \centering
 \includegraphics[width=\linewidth]{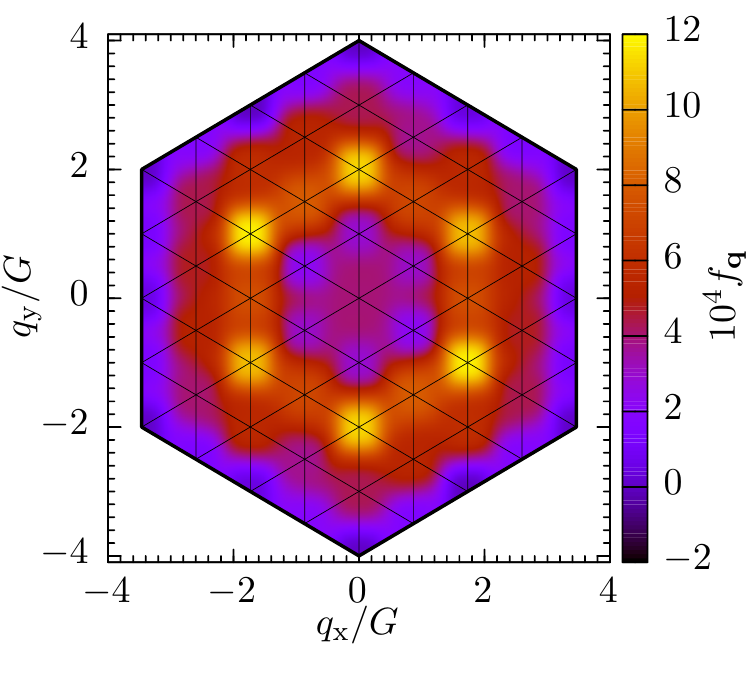}
 \caption{3:1}
 \label{3_1}
\end{subfigure}
\\
\begin{subfigure}{0.85\textwidth}
 \centering
 \includegraphics[width=\linewidth]{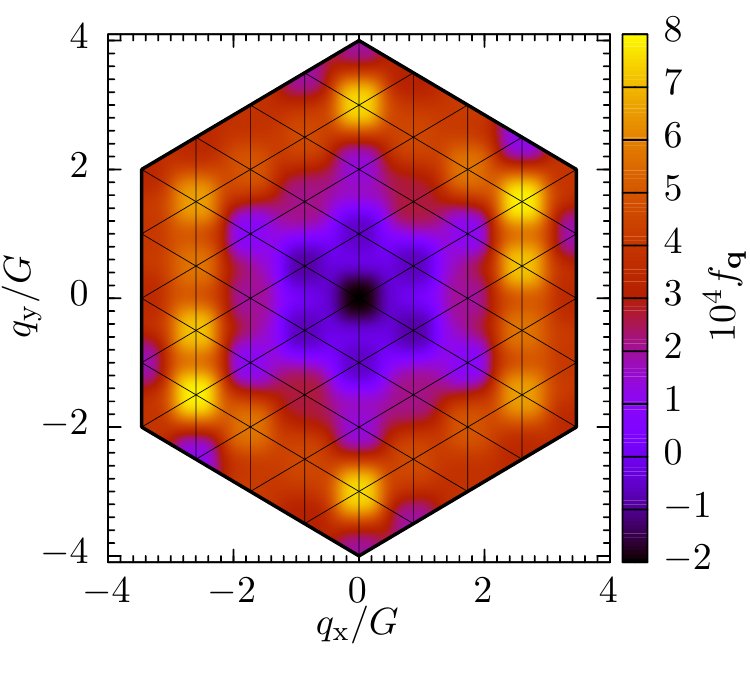}
 \caption{4:1}
 \label{4_1}
\end{subfigure}
\caption{Condensate fraction in momentum space for different ratios of majority to minority species fermions at the Fermi surfaces. 37 majority and 7 minority fermions create a 3:1 spin imbalance at the Fermi surfaces and 61 majority and 7 minority fermions create a 4:1 spin imbalance.}
\label{imbas}
\end{figure}

\subsubsection{Changing spin-imbalance}
Following on from our analysis of the 61 up-spin and 19 down-spin system, we now study two examples of greater spin-imbalance on a triangular lattice shown in Fig.~\ref{imbas}. 37 majority and 7 minority fermions were used to create a 3:1 ratio at the Fermi surfaces with $a=5.6$ and $\reff=0$. Peaks are clearly structured in a ring between $q=\sqrt{3}G$ and $q=2G$ at 10 sample standard deviations above zero. Similarly, 61 majority and 7 minority fermions were used to create a 4:1 ratio at the Fermi surfaces with $a=6.3$ and $\reff=0$, and the condensate fraction once again forms a ring structure peaked from $q=\sqrt{7}G$ to $q=\sqrt{12}G$ at 7 sample standard deviations above zero. Pairing FFLO peaks cannot be seen, and a BCS peak is even more strongly suppressed than in the 2:1 imbalanced case. These systems both provide further strong evidence of a spatially modulated superconducting order parameter, that is of the communal pairing rather than FFLO phase. A similar characterisation exercise to that described above with comparison to exact diagonalization was conducted on both systems and the communal state indices determined. The system with a 3:1 ratio is most closely described by a $(\nup,\ndo)=(3,1)$ state and that with a 4:1 ratio by a $(\nup,\ndo)=(4,1)$ state.

\begin{figure}[t]
\centering
\includegraphics[width=0.9\linewidth]{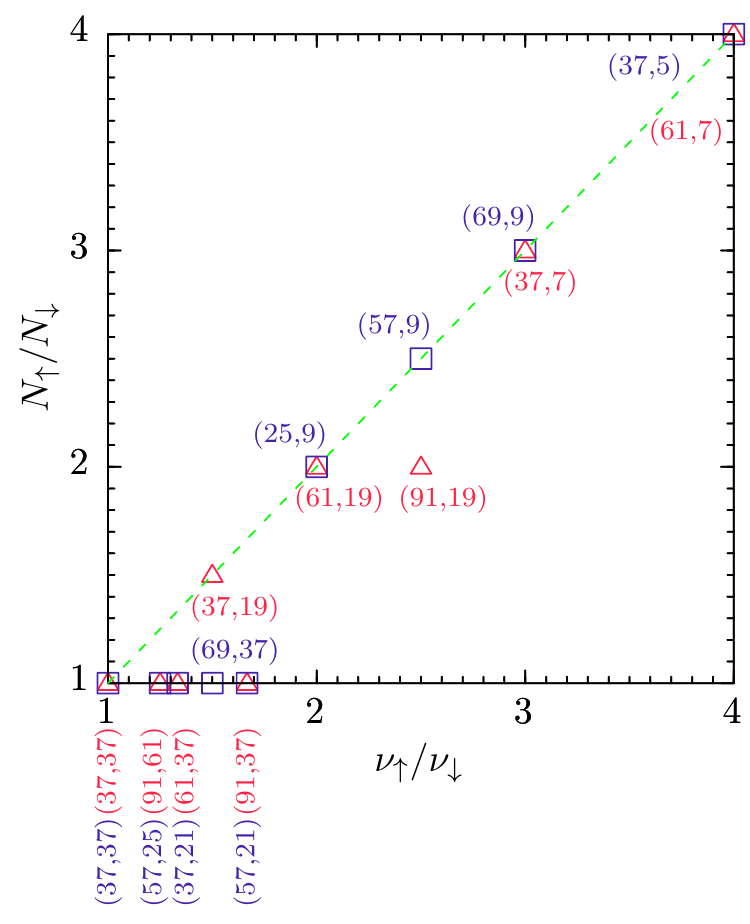}
\caption{Summary plot of the ratio of communal state indices $\nup/\ndo$ to the ratio of densities of states at the respective Fermi surfaces, $\nuup/\nudn$. The red triangles indicate data taken on triangular momentum space lattices and the blue squares indicate data taken on square momentum space lattices. The bracketed number pairs indicate particle numbers for selected systems, with the colour corresponding to the respective lattice types. The line $\nup/\ndo=\nuup/\nudn$ is indicated green.}
\label{nvsnu}
\end{figure}

\subsubsection{Relationship between $\nup/\ndo$ and $\nuup/\nudn$}
We have shown the results from four exemplar systems in detail to demonstrate the emergence of spatially modulated pairing and provided evidence that the form of the spatial modulation observed is characteristic of communal pairing. The analysis was also repeated for 14 systems with other Fermi surface ratios, for both the triangular and square lattices and at different system sizes, and the results analysed to ascertain the communal state indices, $\nup$ and $\ndo$. A summary of the 18 sets of results obtained is shown in \figref{nvsnu}.

The relationship between the ratio of commmunal state indices $\nup/\ndo$ and Fermi surface density of states ratios $\nuup/\nudn$ is well described by the line $\nup/\ndo=\nuup/\nudn$, providing strong evidence for the communal pairing~\cite{multipartsupconfew,commupair} over FFLO. The relationship is particularly strong when the ratio can be written containing small integers \cite{multipartsupconfew} to minimise the product $\nup\ndo$, mitigating the energy penalty for states with high $\nup\ndo$~\cite{commupair}. The correlation coefficient between the gathered data and the line $\nup/\ndo=\nuup/\nudn$ is $R^2=0.95$.

\section{Conclusions}\label{sec:conclusion}
We have observed a spatially modulated superconducting state using DMC. Furthermore, the state is qualitatively different from an FFLO state~\cite{ffloff,fflolo}, having condensate fraction peaks at multiple momenta, as opposed to the single peak expected for FFLO. Exact diagonalisation studies provide corroborating evidence that the distribution of condensate fraction with momenta is more indicative of communal superconductivity~\cite{multipartsupconfew,commupair} than traditional FFLO or crystalline FFLO superconductivity~\cite{fflorevision1,fflorevision2}. We have confirmed convergence of the state with respect to choices of system size, scattering length, pseudopotential cutoff length, DMC timestep, and DMC walker population.

This numerical evidence that builds on previous analytical work~\cite{multipartsupconfew,commupair} provides an interesting challenge for experiments to observe the communal state in physical systems. In real space the superconducting order parameter will exhibit a beat pattern due to the interference between similar $q$-vectors, which could allow the identification of the particular $q$-vectors in the superconductor.  The order parameter and its spread in momentum could be determined in an ultracold atomic gas experiment through density-density correlations measured from time-of-flight experiments~\cite{Altman04}. In contrast, FFLO and crystalline FFLO theories predict sharp peaks in the condensate fraction, as in spin-balanced BCS theory, at fixed magnitude of the pairing momenta.

Additionally, as the communal number pair $(\nup,\ndo)$ is a function of the spin imbalance, multiple phase transitions through several superconducting phases should be observed as the imbalance is increased. Each transition is expected to be second order, and so the communal superconducting phase would be characterized by a series of singularities in the heat capacity and the compressibility, which should be directly observable in ultracold atomic gases~\cite{Ku12} as the spin-imbalance is changed. No such phase transitions are expected for the FFLO phase at fixed temperature.

An orthogonal line of questioning that may be of concern in real experiments is on the possible effects a nonzero effective range might have on the obtained results. Previous work~\cite{reff} suggests that the obtained energy differences from the noninteracting state should increase towards zero, the condensate fraction should be constant over a wide range of scattering lengths, and the momentum density should become more sharply step-like.

Finally, the match with exact diagonalisation studies provides evidence that the elementary excitations above the proposed ground state are well-described by the few fermion analysis~\cite{multipartsupconfew}. This should have novel consequences especially concerning Andreev reflection experiments as the strong correlations between a group of fermions held in a communal state should result in multiple retroreflected holes for a single incident fermion, in sharp contrast to the single hole per fermion expected in normal FFLO theory.

\section*{Acknowledgements}
Data used for this Letter are available online~\cite{dspacefooconduit}. The authors thank Pablo L\'{o}pez R\'{i}os and Thomas Whitehead for useful discussions, and acknowledge the financial support of the NUS and the Royal Society.


\begin{thebibliography}{0}

\bibitem{bcs}
J.~Bardeen, L.~N.~Cooper, and J.~R.~Schrieffer, Phys.~Rev.~{\bf 106}, 162 (1957).

\bibitem{supconelemrev}
J.~Eisenstein, Rev.~Mod.~Phys.~{\bf 26}, 277 (1954).

\bibitem{supconcomprev}
G.~W.~Webb, F.~Marsiglio and J.~E.~Hirsch, Physica C~{\bf 514}, 17 (2015).

\bibitem{consupconquacrys}
R.~N.~Ara\'{u}jo and E.~C.~Andrade, Phys.~Rev.~B~{\bf 100}, 014510 (2019).

\bibitem{consupconhight}
M.~I.~Eremets and A.~P.~Drozdov, Phys.-Usp~{\bf 59}, 1154 (2016).

\bibitem{chandrasekhar_limit}
B.~S.~Chandrasekhar, Appl.~Phys.~Lett.~{\bf 1}, 7 (1962).

\bibitem{clogston_limit}
A.~M.~Clogston, Phys.~Rev.~Lett.~{\bf 9}, 266 (1962).

\bibitem{breachsupercon1}
E.~Gubankova, W.~V.~Liu, and F.~Wilczek, Phys.~Rev.~Lett.~{\bf 91}, 032001 (2003).
\bibitem{breachsupercon2}
W.~V.~Liu and F.~Wilczek, Phys.~Rev.~Lett.~{\bf 90}, 047002 (2003).

\bibitem{breachsupercon3}
M.~M.~Forbes, E.~Gubankova, W.~V.~Liu, and F.~Wilczek, Phys.~Rev.~Lett.~{\bf 94}, 017001 (2005).

\bibitem{ffloff}
P.~Fulde and R.~A.~Ferrell, Phys.~Rev.~{\bf 135}, A550 (1964).

\bibitem{fflolo}
A.~I.~Larkin, Y.~N.~Ovchinnikov, Sov.~Phys.~JETP~{\bf 20}, 762 (1965).

\bibitem{multipartsupconfew}
T.~M.~Whitehead and G.~J.~Conduit, Phys.~Rev.~B~{\bf 97}, 014502 (2018).

\bibitem{commupair}
D.~C.~W.~Foo, T.~M.~Whitehead and G.~J.~Conduit, EPL~{\bf 126}, 6 (2019).

\bibitem{Onnes1}
H.~K.~Onnes, Comm.~Phys.~Lab.~Univ.~Leiden~{\bf 120b}, (1911).

\bibitem{Onnes2}
H.~K.~Onnes, Comm.~Phys.~Lab.~Univ.~Leiden~{\bf 122b}, (1911).

\bibitem{ffloexpthf}
Y.~Matsuda and H.~Shimahara, J.~Phys.~Soc.~Jpn.~{\bf 76}, 051005 (2007).

\bibitem{ffloexptfe1}
D.~A.~Zocco, K.~Grube, F.~Eilers, T.~Wolf and H.~v.~L\"{o}hneysen, Phys.~Rev.~Lett.~{\bf 111}, 057007 (2013).

\bibitem{ffloexptfe2}
C.~Cho, J.~H.~Yang, N.~F.~Q.~Yuan, J.~Shen, T.~Wolf and R.~Lortz, Phys.~Rev.~Lett.~{\bf 119}, 217002 (2017).

\bibitem{ffloexptfe3}
M.~Nikolo, J.~Singleton, D.~Solenov, J.~Jiang, J.~D.~Weiss and E.~E.~Hellstrom, J.~Supercond.~Nov.~Magn.~{\bf 30}, 561 (2017).

\bibitem{ffloexptfe4}
T.~Terashima, N.~Kikugawa, A.~Kiswandhi, E.~Choi, K.~Kihou, S.~Ishida, C.~Lee, A.~Iyo, H.~Eisaki and S.~Uji, Phys.~Rev.~B~{\bf 99}, 094508 (2019).

\bibitem{ffloexptdw}
L.~Zhang, Y.~Ge, X.~Shang and Y.~Gao, Mod.~Phys.~Lett.~B~{\bf 33}, 1950082 (2019).

\bibitem{ffloexptorg1}
R.~Lortz, Y.~Wang, A.~Demuer, P.~H.~M.~B\"{o}ttger, B.~Bergk, G.~Zwicknagl, Y.~Nakazawa, and J.~Wosnitza, Phys.~Rev.~Lett.~{\bf 99}, 187002 (2007).

\bibitem{ffloexptorg2}
R.~Beyer, B.~Bergk, S.~Yasin, J.~A.~Schlueter and J.~Wosnitza, Phys.~Rev.~Lett.~{\bf 109}, 027003 (2012).

\bibitem{ffloexptorg3}
G.~Koutroulakis, H.~K\"{u}hne, J.~A.~Schlueter, J.~Wosnitza and S.~E.~Brown, Phys.~Rev.~Lett.~{\bf 116}, 067003 (2016).

\bibitem{ffloexptorg4}
S.~Sugiura, T.~Isono, T.~Terashima, S.~Yasuzuka, J.~A.~Schlueter and S.~Uji, npj~Quantum~Mat.~{\bf 4}, 7 (2019).

\bibitem{ffloexptsflayer}
S.~V.~Mironov, D.~Y.~Vodolazov, Y.~Yerin, A.~V.~Samokhvalov, A.~S.~Mel'nikov and A.~Buzdin, Phys.~Rev.~Lett.~{\bf 121}, 077002 (2018).

\bibitem{ffloquasicrys}
S.~Sakai and R.~Arita, Phys.~Rev.~Research~{\bf 1}, 022002(R) (2019).

\bibitem{ffloexptopt1}
M.~W.~Zwierlein, A.~Schirotzek, C.~H.~Schunck, and W.~Ketterle, Science~{\bf 311} 492 (2006).

\bibitem{ffloexptopt2}
Y.~A.~Liao, A.~S.~C.~Rittner, T.~Paprotta, W.~Li, G.~B.~Partridge, R.~G.~Hulet, S.~K.~Baur and E.~J.~Mueller, Nature~{\bf 467} 567 (2010).

\bibitem{ffloorg}
F.~Piazza, W.~Zwerger and P.~Strack, Phys.~Rev.~B~{\bf 93} 085112 (2016).

\bibitem{lowd}
A.~Ptok, J.~Phys.:~Condens.~Matter~{\bf 29} 475901 (2017).

\bibitem{uniformtrap}
B.~Mukherjee, Z.~Yan, P.~B.~Patel, Z.~Hadzibabic, T.~Yefsah, J.~Struck and M.~W.~Zwierlein, Phys.~Rev.~Lett.~{\bf 118}, 123401 (2017).

\bibitem{trapfg1}
B.~Mukherjee, P.~B.~Patel, Z.~Yan, R.~J.~Fletcher, J.~Struck and M.~W.~Zwierlein, Phys.~Rev.~Lett.~{\bf 122}, 203402 (2019).

\bibitem{trapfg2}
Z.~Yan, P.~B.~Patel, B.~Mukherjee, R.~J.~Fletcher, J.~Struck and M.~W.~Zwierlein, Phys.~Rev.~Lett.~{\bf 122}, 093401 (2019).

\bibitem{lah}
M.~Somayazulu, M.~Ahart, A.~K.~Mishra, Z.~M.~Geballe, M.~Baldini, Y.~Meng, V.~V.~Struzhkin and R.~J.~Hemley, Phys.~Rev.~Lett.~{\bf 122} 027001 (2019).

\bibitem{magnetar}
T.~Lee, R.~Yoshiike and T.~Tatsumi, JPS~Conf.~Proc.~{\bf 20} 011006 (2018).

\bibitem{quark}
T.~Tatsumi, JPS~Conf.~Proc.~{\bf 20} 011008 (2018).

\bibitem{casalbuoni}
R.~Casalbuoni and G.~Nardulli, Rev.~Mod.~Phys.~{\bf 76} 263 (2004).
\bibitem{fflostability1}
G.~Orso, Phys.~Rev.~Lett.~{\bf 98}, 070402 (2007).

\bibitem{fflostability2}
H.~Hu, X.~J.~Liu and P.~D.~Drummond, Phys.~Rev.~Lett.~{\bf 98}, 070403 (2007).

\bibitem{fflostability3}
M.~M.~Parish, S.~K.~Baur, E.~J.~Mueller and D.~A.~Huse, Phys.~Rev.~Lett.~{\bf 99}, 250403 (2007).

\bibitem{riospairingorb}
R.~Maezono, P.~L\'{o}pez~R\'{i}os, T.~Ogawa and R.~J.~Needs, Phys.~Rev.~Lett.~{\bf 110}, 216407 (2013).

\bibitem{dmcfermion1}
L.~K.~Wagner and D.~M.~Ceperley, Rep.~Prog.~Phys.~{\bf 79}, 094501 (2016).

\bibitem{dmcfermion2}
A.~Galea, H.~Dawkins, S.~Gandolfi and A.~Gezerlis, Phys.~Rev.~A~{\bf 93}, 023602 (2016).

\bibitem{casino}
R.~Needs, M.~Towler, N.~Drummond and P.~L\'{o}pez~R\'{i}os, \textsc{casino} User's Guide Version 2.13 (2015).

\bibitem{2dscapseudopot}
T.~M.~Whitehead, L.~M.~Schonenberg, N.~Kongsuwan, R.~J.~Needs and G.~J.~Conduit, Phys.~Rev.~A {\bf 93}, 042702 (2016).

\bibitem{drumjas}
N.~D.~Drummond, M.~D.~Towler and R.~J.~Needs, Phys.~Rev.~B {\bf 70}, 235119 (2004).

\bibitem{nujas}
T.~M.~Whitehead, M.~H.~Michael and G.~J.~Conduit, Phys.~Rev.~B {\bf 94}, 035157 (2016).

\bibitem{qmcmethods}
W.~M.~C.~Foulkes, L.~Mitas, R.~J.~Needs and G.~Rajagopal, Rev.~Mod.~Phys. {\bf 73}, 33 (2001).
\bibitem{qmcreview}
R.~J.~Needs, M.~D.~Towler, N.~D.~Drummond and P.~L\'{o}pez~R\'{i}os, J.~Phys.~Condens.~Matter {\bf 22}, 023201 (2010).

\bibitem{dmcintro}
I.~Kosztin, B.~Faber and K.~Schulten, Am.~J.~Phys. {\bf 64}, 633 (1996).

\bibitem{finitesizeeffect}
N.~D.~Drummond, R.~J.~Needs, A.~Sorouri and W.~M.~C.~Foulkes, Phys.~Rev.~B {\bf 78}, 125106 (2008).

\bibitem{simdark}
D.~Frenkel, Eur.~Phys.~J.~Plus  128:~10(2013).

\bibitem{dmcN1}
A.~Galea, H.~Dawkins, S.~Gandolfi, and A.~Gezerlis, Phys.~Rev.~A {\bf 93}, 023602 (2016).

\bibitem{dmcN2}
G.~E.~Astrakharchik, J.~Boronat, J.~Casulleras, and S.~Giorgini, Phys.~Rev.~Lett {\bf 93}, 200404 (2004).

\bibitem{2dscaverhaar84}
B.~J.~Verhaar, J.~P.~H.~W.~van den Eijnde, M.~A.~J.~Voermans and M.~M.~J.~Schaffrath, J.~Phys.~A: Math.~Gen.~{\bf 17}, 595 (1984)

\bibitem{fermigasreview}
W.~Ketterle and M.~W.~Zwierlein, Rivista Del Nuovo Cimento {\bf 31}, 5 (2008).

\bibitem{shi}
H.~Shi, S.~Chiesa and S.~Zhang, Phys.~Rev.~A {\bf 92}, 033603 (2015).

\bibitem{fflorevision3}
A.~Bulgac, M.~M.~Forbes and A.~Schwenk, Phys.~Rev.~A~{\bf 78}, 033607 (2008).

\bibitem{fflo1d}
M.~Casula and D.~M.~Ceperley, Phys.~Rev.~Lett.~{\bf 97}, 020402 (2006).

\bibitem{fewpart}
P.~O.~Bugnion, J.~A.~Lofthouse and G.~J.~Conduit, Phys.~Rev.~Lett.~{\bf 111}, 045301 (2013).

\bibitem{fflorevision1}
H.~Shimahara, J.~Phys.~Soc.~Japan~{\bf 67}, 736 (1998).

\bibitem{fflorevision2}
C.~Mora and R.~Combescot, EPL~{\bf 66}, 833 (2004).

\bibitem{Altman04}
E.~Altman, E.~Demler, and M.~D.~Lukin, Phys.~Rev.~A~{\bf 70}, 013603 (2004).

\bibitem{Ku12}
M.~J.~H.~Ku, A.~T.~Sommer, L.~W.~Cheuk, and M.~W.~Zwierlein, Science~{\bf 335}, 563
(2012).

\bibitem{reff}
L.~M.~Schonenbergm P.~C.~Verpoort and G.~J.~Conduit, Phys.~Rev.~A~{\bf 96} 023619 (2017).

\bibitem{dspacefooconduit}
D.~C.~W.~Foo and G.~J.~Conduit, University of Cambridge DSpace repository www.openaccess.cam.ac.uk.

\bibitem{fflo3dphasediag1}
D.~E.~Sheehy and L.~Radzihovsky, Phys.~Rev.~Lett.~{\bf 96}, 060401 (2006).

\bibitem{fflo3dphasediag2}
D.~E.~Sheehy and L.~Radzihovsky, Ann.~Phys.~{\bf 322}, 1790 (2007).
\bibitem{qmcaccuracy}
W.~D.~Parker, J.~W.~Wilkins, R.~G.~Hennig, Phys.~Status~Solidi~B~{\bf 248}, 2 (2011).

\bibitem{qmcefficiency}
R.~M.~Lee, G.~J.~Conduit, N.~Nemec, P.~L\'{o}pez~R\'{i}os, and N.~D.~Drummond, Phys.~Rev.~E~{\bf 83}, 066706 (2011).

\bibitem{qmctimestep}
J.~Vrbik and S.~M.~Rothstein, Int.~J.~Quantum~Chem.~{\bf 29}, 461 (1986).




\end{thebibliography}

\begin{appendix}
\appendix
\section{Simulation parameters}\label{subsec:simparams}
It is of essential importance in any numerical study that the underlying distribution sampled from is well-behaved to ensure applicabaility of statistical measures such as sample variance. Here, histograms of the accumulated energy values from up to $10^7$ samples did not reveal any evidence of non-normal behaviour and so the sample error is taken to be a good estimate of the true statistical error. We now explore the robustness of our conclusions against the choice of simulation parameters, specifically the scattering length, the pseudopotential cutoff length, the DMC timestep, and the number of DMC walkers. The system was selected to have 61 majority and 19 minority spin fermions.

\begin{figure}[t]
	\centering
	\includegraphics[width=0.9\linewidth]{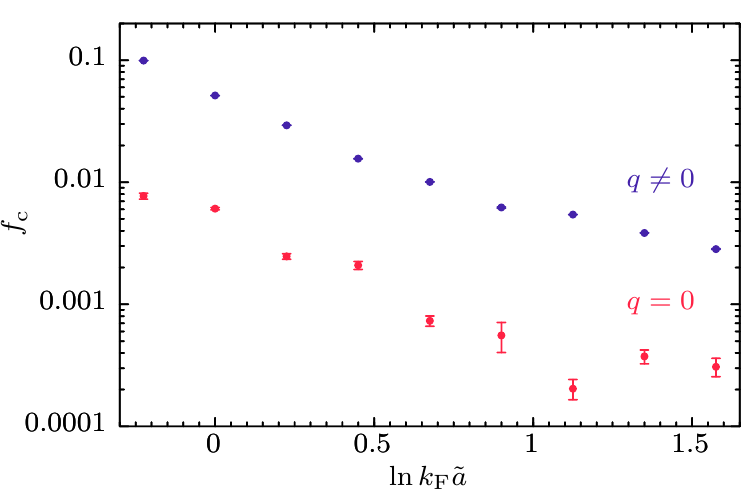}
	\caption{Fraction of total condensate fraction $f_\mathrm{c}$ contained in the communal state (blue) and BCS (red) peaks against inverse interaction strength $\ln\kF\tilde{a}$.}
	\label{fcvsloga}
\end{figure}

\paragraph{Scattering length}
The condensate fraction was found to be robust against large variation of the scattering length and therefore of the interaction strength, with the ratio of the condensate fraction held in the peaks of a given $q$ to that of the total condensate fraction summed over all $q$, $f_c$, holding roughly constant, despite the raw values of the condensate fraction getting larger at higher interaction strengths. This is shown in \figref{fcvsloga}, where the constant $f_c$ indicates that the region of stability for the states of finite $q$ in 2D is much wider than corresponding region in the 3D phase diagram \cite{fflo3dphasediag1,fflo3dphasediag2}.

\begin{figure}[b]
\centering
\includegraphics[width=0.9\linewidth]{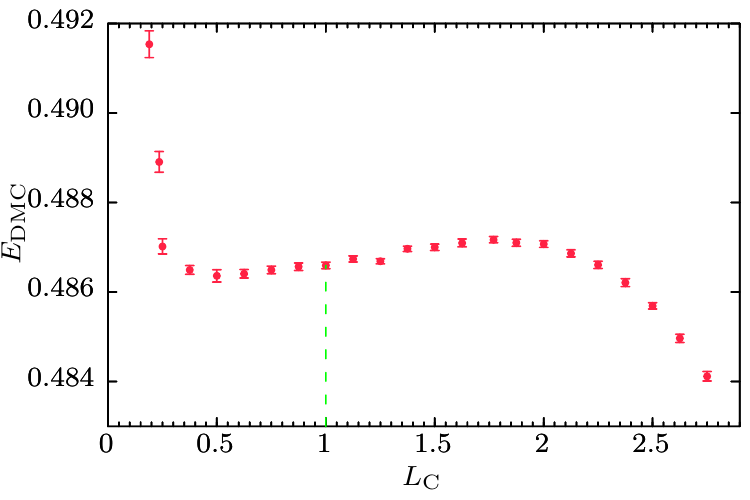}
\caption{Graph of DMC energy, $E_{\mathrm{DMC}}$, against cutoff length $\cutlen$ of the pseudopotential. The energy is shown to vary considerably with cutoff length.}
\label{cutofflength}
\end{figure}

\paragraph{Pseudopotential cutoff length}The effect of altering the cutoff length $\cutlen$ of the UTP on the DMC energy was investigated with results shown in Fig.~\ref{cutofflength}. In contrast to theoretical predictions, the DMC energy, $E_{\mathrm{DMC}}$, is shown to vary considerably with $\cutlen$, with high $E_{\mathrm{DMC}}$ at low $\cutlen$ and vice versa, with an intermediate plateau. All data were gathered with a trial wavefunction that had the same number of variational parameters in the pairing orbital and the Jastrow factor, optimised for the specific potentials, with all other simulation parameters constant. 

The high $E_{\mathrm{DMC}}$ values obtained for low $\cutlen$ are primarily due to poor fit of wavefunction; as a low $\cutlen$ leads to a deep and rapidly varying UTP over a smaller region of space, the trial wavefunction should also include higher order terms to reflect the rapid variation of the UTP. Limiting the number of variational parameters in order to make the results more easily comparable thus leads to a poorer fit of wavefunction as $\cutlen$ decreases, resulting in the ground state not being adequately projected by DMC and increasing the energy. In addition, the deep, rapidly varying UTP results in a greater spread of values for the local energy, leading to a higher sample variance. For $\cutlen = 0.125\rs$ (not in figure), the variation in local energies was wild enough that it eventually lead to extinction of all walkers through the DMC branching factor, and as such no data could be gathered under the simulation parameters selected for all other values of $\cutlen$. 

The low $E_{\mathrm{DMC}}$ values obtained for high $\cutlen$ in contrast are due to higher order interactions beyond pairing, as the UTP now extends over a large enough region that the formation of larger correlated structures is possible. These higher order interactions further decrease the energy and indicate a breakdown of the UTP's ability to emulate a contact interaction, which should only result in pair point interactions for a reasonable fermion density. 

It is desirable to have an easily optimised wavefunction with a low variance and no evidence of three body effects. Therefore, an intermediate value of $\cutlen = \rs$ was chosen for all other tests and simulations.

\begin{figure}[t]
\centering
\includegraphics[width=0.9\linewidth]{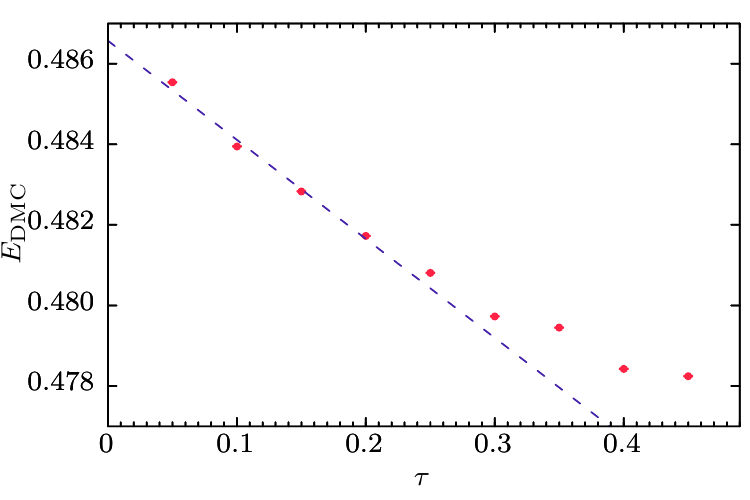}
\caption{Graph of DMC energy $E_{\mathrm{DMC}}$ against DMC timestep $\tau$. Raw data is presented in red with a best fit line for the linear regime in dashed blue.}
\label{evstau}
\end{figure}

\begin{figure}[t]
\centering
\includegraphics[width=0.9\linewidth]{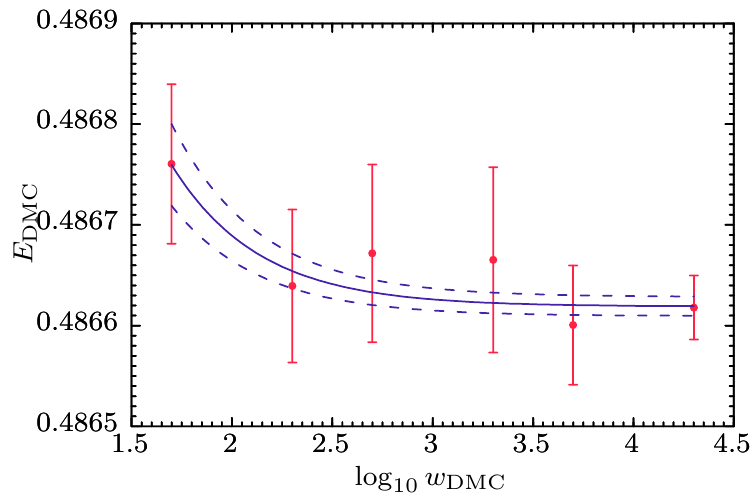}
\caption{Graph of DMC energy $E_{\mathrm{DMC}}$ against the Log of the number of DMC walkers $W_{\mathrm{DMC}}$ for a fixed product of the number of walkers and number of steps, $W_{\mathrm{DMC}}N=2.5\times 10^{7}$.}
\label{evlnn}
\end{figure}

\paragraph{DMC timestep}The DMC algorithm is only exact in the limit of zero timestep $\tau$. However, the computational effort required to achieve a given error bar scales as $1/\tau$, so it is not feasible to simply use infinitesimally small timesteps. For sufficiently small $\tau$, the DMC energy varies linearly with the timestep, $E_{\mathrm{DMC}}(\tau) = E_0 + \kappa\tau$ where $E_0$ is the true ground state energy. Hence, if the linear regime can be identified, it is possible to extrapolate the DMC results down to zero timestep, and efficient algorithms have been proposed for this~\cite{qmcaccuracy,qmcefficiency,qmctimestep}.

We follow the algorithm in ~\cite{qmcefficiency} to extrapolate to zero timestep using the results shown in Fig.~\ref{evstau}. Taking the maximum timestep of the linear regime to be $\tau_2 = 0.20$, we set $\tau_1 = \tau_2/4$, and use a total number of steps $T_1 = 2.5\times 10^7$ and  $T_2 = T_1/8 = 3.125\times 10^6$ respectively to obtain an energy of 0.48684(2).

\paragraph{DMC walkers}The DMC algorithm makes use of the drift-diffusion of a regulated number of walkers for a specified amount of time to obtain expectation values of physical observables. The total computation time $T$ therefore is a function of not only the time-averaged number of DMC walkers, $W_{\mathrm{DMC}}$, but also of the number of timesteps, $N$, as $T=W_{\mathrm{DMC}}N$. The effect of varying $W_{\mathrm{DMC}}$ while keeping $T$ constant was investigated and the results shown in Fig.~\ref{evlnn}. The DMC energy does not vary significantly even as $W_{\mathrm{DMC}}$ spans several orders of magnitude while the sample variance decreases for $W_{\mathrm{DMC}} > 2000$. It is thus preferable to have a high number of walkers propogated a few steps forward in imaginary time than to have a small number of walkers propogate for a long imaginary time.

\end{appendix}
\end{document}